\documentclass[fleqn,twoside]{article}
\usepackage[headings]{espcrc2}

\readRCS
$Id: espcrc2.tex,v 1.2 2004/02/24 11:22:11 spepping Exp $
\usepackage{graphicx}

\newcommand{\AmS}{{\protect\the\textfont2
  A\kern-.1667em\lower.5ex\hbox{M}\kern-.125emS}}

\title{Photon-Induced Physics with Heavy-Ion Beams in ALICE}

\author{Joakim Nystrand\address{Department of Physics and Technology, 
University of Bergen, Bergen, Norway} for the ALICE Collaboration}
       
\runtitle{2-column format camera-ready paper in \LaTeX}
\runauthor{J. Nystrand}
\runtitle{Photon-Induced Physics with Heavy-Ion Beams in ALICE}

\begin{document}

\begin{abstract}
The possibilities to study ultra-peripheral collisions, in particular 
exclusive vector meson production, with the ALICE detector is reviewed. 
\vspace{1pc}
\end{abstract}

\maketitle

Photon-induced interactions have traditionally been studied with lepton beams in fixed target 
experiments and at colliders. But two--photon and photon--hadron interactions can occur also 
when the lepton beams are replaced by high energy protons or heavy 
ions\cite{Baltz:2007kq,Bertulani:2005ru}. The electromagnetic interactions can be studied in 
ultra-peripheral collisions where the 
impact parameters are larger than the sum of the projectile radii and no hadronic interactions 
occur. 
This paper will discuss the possibilities for studying ultra-peripheral collisions in the 
ALICE experiment.

\section{The ALICE Experiment}

ALICE is an experiment at the CERN Large Hadron Collider (LHC) primarily designed to study 
the physics of strongly interacting matter in nucleus-nucleus 
collisions\cite{Carminati:2004fp,Alessandro:2006yt}. The ALICE detector consists of a 
central barrel, a forward muon arm, and a set of smaller, forward detectors. 

The central barrel, which covers the pseudorapidity range $| \eta | <$~0.9, has a charged particle 
tracking system consisting of an Inner Tracking System (ITS) in combination with a cylindrical 
Time-Projection 
Chamber (TPC). Particle identification is obtained from the energy loss in the tracking detectors 
in combination with information from a Time-of-Flight (TOF) detector, a Transition Radiation Detector 
(TRD), and an imaging Cherenkov detector (HMPID). In addition, the central barrel contains a 
high resolution Photon Spectrometer (PHOS) and an Electromagnetic Calorimeter (EMCAL).  
The HMPID, PHOS and EMCAL cover only parts of the central barrel acceptance. 

The central barrel is designed to be able to handle multiplicities up to $dn_{ch}/d\eta =$~8000, 
a conservative estimate of the maximum multiplicities expected in central Pb+Pb collisions at the 
LHC. All charged tracks with transverse momenta ($p_T$) greater than $0.2$~GeV/c can be reconstructed, 
with a momentum resolution changing from about 0.7\% at $p_T = 1$~GeV/c to 3.5\% at $p_T = 100$~GeV/c. 

The muon arm covers the pseudorapidity range -2.5~$> \eta >$~-4.0 and consists of an arrangement of 
absorbers, a large dipole magnet, ten planes of cathode pad tracking chambers, and four planes of 
resistive-plate triggering chambers. The muon arm is capable of reconstructing $J/\Psi$ and 
$\Upsilon$ vector mesons through their di--muon decay channel with transverse momenta down to 
$p_T =$~0 and with a resolution in invariant mass of 70 and 100~MeV/c$^2$, respectively. 


Several smaller detectors are located at forward angles. These are the photon multiplicity 
detector (PMD), the forward multiplicity detector (FMD), the T0 and V0 detectors, and the 
Zero-Degree calorimeters (ZDC). The T0, Cherenkov radiation detectors, and V0, plastic scintillators,  
cover about 0.5 and 2.0 units of pseudorapidity, respectively, on each side of the 
interaction point. 

Signals from the V0, T0, ITS (Si-Pixel), TOF, TRD, PHOS, EMCAL, and the muon trigger chambers 
are available in the lowest ALICE trigger level (Level~0). 
The ZDCs are located too far from the interaction point to be included in Level~0, but the 
information is available in the higher trigger levels. 

An outline of the ALICE detector is shown in Fig.~1. 

\begin{figure*}[htb]
\vspace{9pt}
\begin{center}
\includegraphics*[width=11cm]{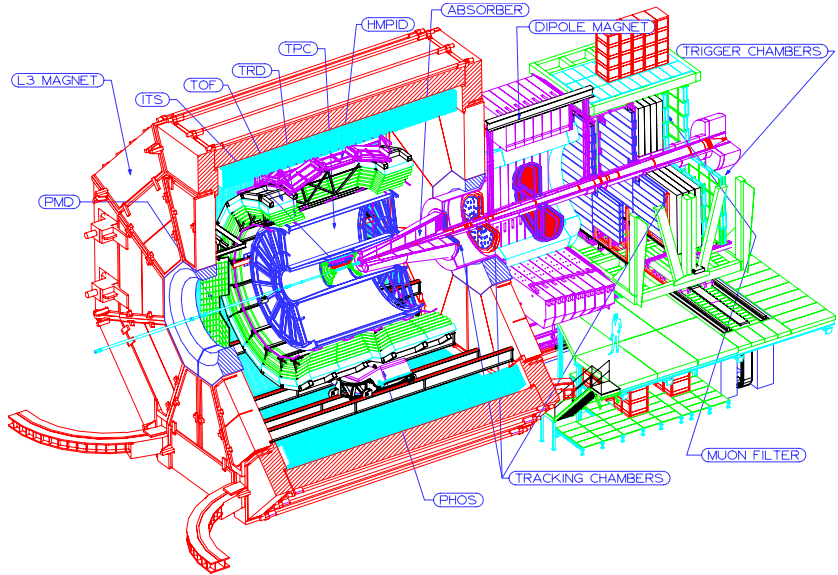}
\end{center}
\caption{Layout of the ALICE detector.  The central barrel is embedded 
in the L3 Magnet, which provides a 0.5~T magnetic field. The magnet has an outer radius of 7.9~m and an 
overall length of 14~m, and the TPC has an outer radius of 250~cm, and a length of 500~cm.}
\label{fig:alice}
\end{figure*}

\section{Ultra-Peripheral Collisions in ALICE} 

Ultra-peripheral collisions will have a very different topology compared with the 
central, hadronic interactions, which are the main focus of the ALICE experiment. 
The versatility of the ALICE detector should allow these interactions to be studied, however, 
but they will require different trigger and analysis techniques. Ultra-peripheral collisions 
can be studied both with heavy-ion and proton beams. 

The photons from the electromagnetic field of one of the nuclei (or protons) can interact with the 
other nucleus in several ways. The interaction can be purely electromagnetic 
(two-photon interaction) or the photon can interact hadronically with the target nucleus 
(photonuclear interaction). The photonuclear interactions can either lead to the target nucleus 
breaking up (``inelastic'') or remaining intact (``elastic''). The photonuclear 
interactions can be further subdivided depending on if the photon first fluctuates to a hadronic 
state (resolved interactions) or if it interacts as a bare photon (direct interactions). 

Two-photon and elastic photonuclear interactions are characterized by two rapidity gaps void of 
particles on both side of the produced system. 
If the fields couple coherently to the entire nucleus, the total 
transverse momentum will be determined by the nuclear form factors and will typically be 
small compared with that for competing processes. 
The inelastic photonuclear interactions have a single 
rapidity gap between the photon-emitting nucleus and the produced system.

The cross section for inelastic photonuclear interactions is dominated by resolved processes, but 
the cross sections for direct photon-parton interactions are large. For example, the cross 
section to produce a $c \overline{c}$--pair through $\gamma$--gluon fusion in Pb+Pb collisions  
is about 1~barn at the LHC\cite{Klein:2002wm}.The cross section for photon-induced di-jet production is also 
appreciable with, e.g., a cross section larger than 1~$\mu$b for jets with $p_T >$~50~GeV/c and 
rapidity $|y| <$~1 in Pb+Pb collisions\cite{Strikman:2005yv}.    

The elastic photonuclear reactions are dominated by exclusive vector meson 
production\cite{Klein:1999qj}. For comparable final states, the cross sections for two-photon 
production are typically two or three orders of magnitude smaller than for photonuclear 
production. 

The characteristics of ultra-peripheral collisions form the basis for defining the trigger and 
separating the ultra-peripheral events from the hadronic interactions and various types of background 
processes. 

ALICE is designed to handle multiplicities of several thousand particles in a single event. 
The multiplicities 
in ultra-peripheral collisions are normally much lower and reconstructing the events should not pose a 
problem. The key challenge is instead to implement an efficient trigger.  

In central collisions, it is possible to trigger on charged particles emitted outside the acceptance of the 
central barrel, since the produced particles are distributed more or less evenly over the entire rapidity 
interval. The default trigger for hadronic interactions is therefore based on a coincident 
signal in the V0 and T0 detectors on both sides of the interaction point. For ultra-peripheral collisions, 
the idea is instead to let the V0 and T0 
detectors define rapidity gaps by requiring an absence of a signal in those detectors in coincidence with 
a low multiplicity trigger around mid-rapidity. The low multiplicity trigger can be implemented 
through the 
Si-pixel and TOF detectors in the central barrel or through the muon trigger chambers. 
For two-photon and elastic photonuclear interactions, the T0 and V0 detectors on both sides
should be empty, whereas for inelastic photonuclear interactions one should require a signal in the 
V0 and T0 on one side while the other side should be empty. 

ALICE should have the capability to study many different types of ultra-peripheral collisions, 
with and without breakup of the nuclei. The focus so far has, however, been on exclusive production 
where both nuclei remain intact, in particular exclusive vector meson production. This will be discussed 
in more detail for heavy-ion and proton-proton interactions in the following two sections. 

\begin{figure}[tb]
\vspace{9pt}
\begin{center}
\includegraphics*[width=6.5cm]{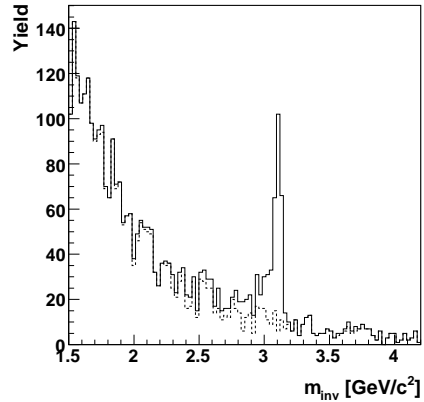}
\end{center}
\caption{Invariant mass distribution of $e^+e^-$--pairs in ultra-peripheral Pb+Pb collisions. The dashed 
histogram shows the continuum contribution from two-photon interactions.}
\label{fig:minvpbpb}
\end{figure}

\section{Exclusive Vector Meson Production in Heavy-Ion Collisions} 

\begin{figure*}[bht]
\vspace{9pt}
\begin{center} 
\includegraphics*[width=5.8cm]{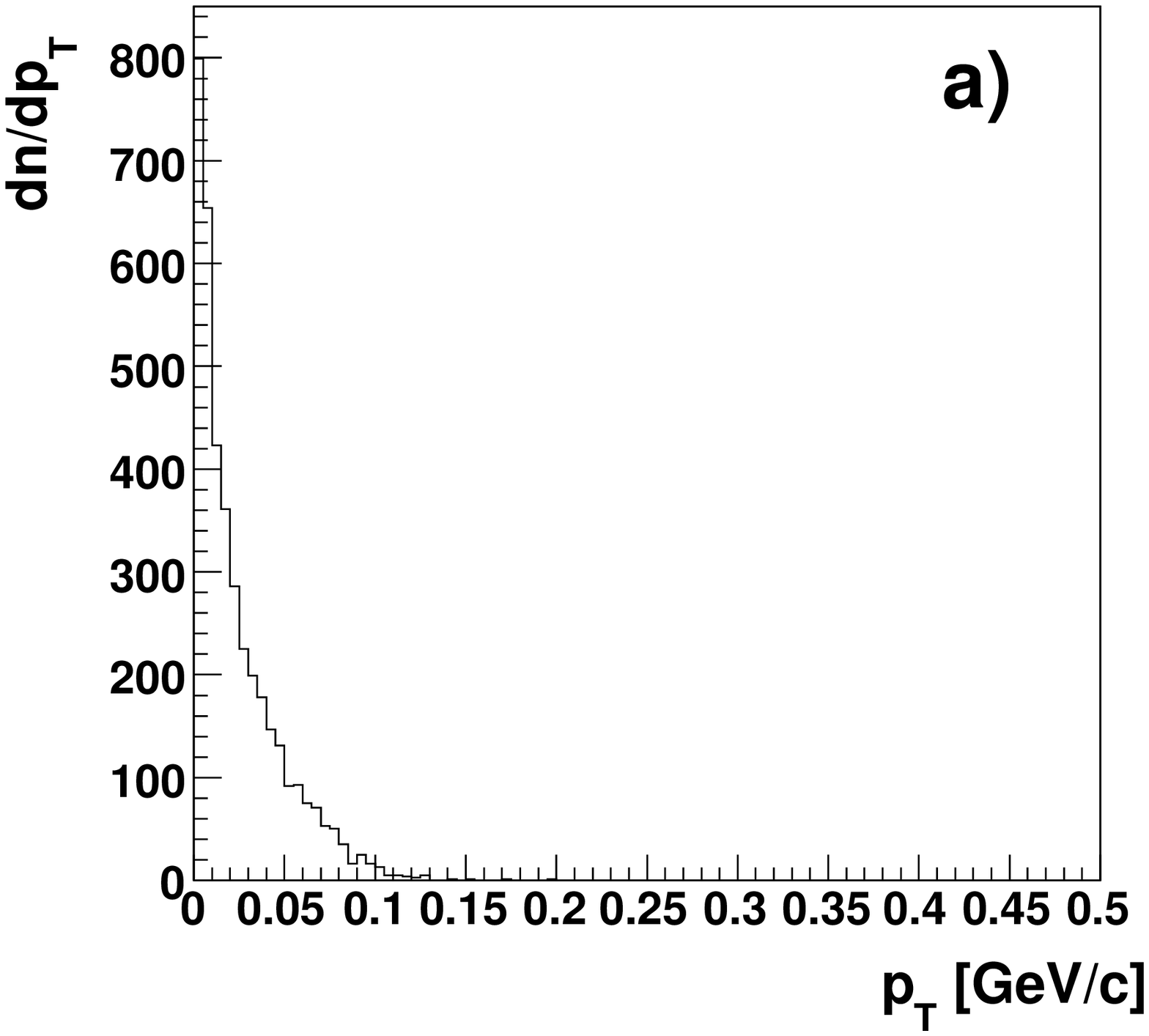}
\includegraphics*[width=5.8cm]{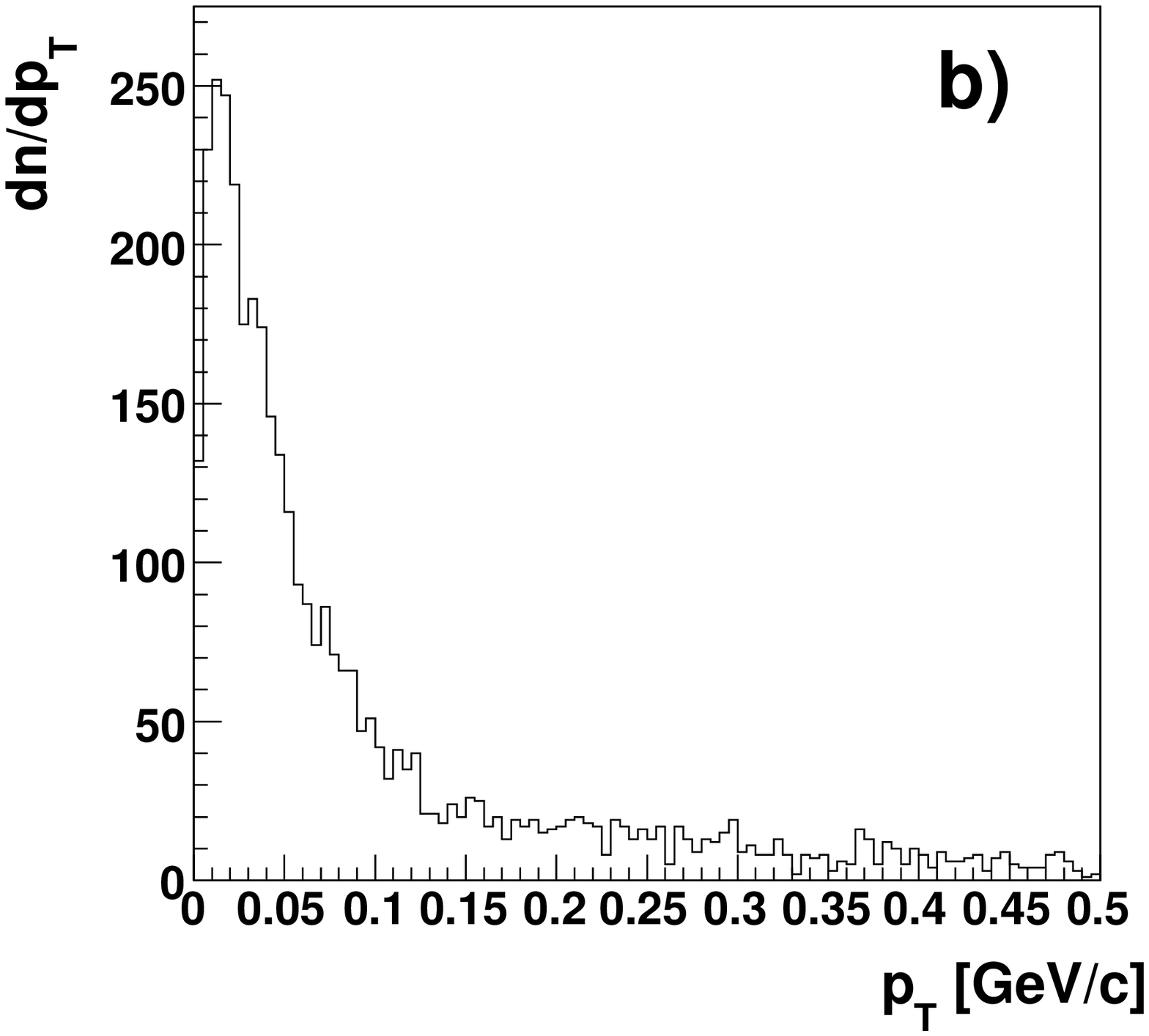}
\end{center} 
\caption{Generated (a) and reconstructed (b) transverse momentum distributions of vector mesons and  
$e^+e^-$--pairs from two-photon interactions in Pb+Pb collisions.}
\label{fig:ptpbpb}
\end{figure*}

At high photon energies and low virtualities, a photon may fluctuate into a vector meson 
and remain in that state for times that are long compared with the time it takes for it 
to pass nuclear distances. 
While in the vector meson state, the photon may scatter diffractively off the 
target nucleus and emerge as a real vector meson. The cross sections for exclusive photoproduction 
of vector mesons are large in heavy-ion interactions at the LHC\cite{Klein:1999qj}. 
Table~1 shows the the cross sections and expected detection rates for $\rho^0$, J/$\Psi$, and $\Upsilon$ 
in ALICE in a one month (10$^6$~s) Pb+Pb run. The detection rates were calculated from the 
geometrical acceptance of the central barrel ($| \eta | <$~0.9, $p_T >$~0.2~GeV/c).  


In this section, the possibilities to reconstruct exclusively produced J/$\Psi$ and $\Psi'$ 
mesons in the ALICE central barrel will be discussed. The J/$\Psi$ and $\Psi'$ are assumed to 
decay via their $e^+e^-$ decay channel. The dominating background process is expected to be 
continuum production of $e^+e^-$--pairs in two-photon interactions\cite{Nystrand:2004vn}.

Other backgrounds include hadronic interactions in peripheral nucleus--nucleus collisions, beam-gas 
interactions, cosmic rays (mainly at the trigger level), and incoherent photonuclear interactions. 
These backgrounds have been studied through simulations\cite{Alessandro:2006yt}. 
The results show that the coherent and exclusive events can be identified with good signal 
to background ratios when the entire event is reconstructed and a cut is applied on the summed  
$p_T$ of the event. This is confirmed by the measurements of exclusive production of $\rho^0$ and 
J/$\Psi$ mesons at RHIC, where similar techniques were used\cite{Adler:2002sc,d'Enterria:2006ep}. 


A simulation has been performed based on 1.5 hours of running at a Pb+Pb luminosity of  
$5 \cdot 10^{26}$~cm$^{-2}$s$^{-1}$. 
Three samples of events were generated: 375,000 
$\gamma \gamma \rightarrow e^+ e^-$ events with an invariant mass of the $e^+e^-$--pairs 
greater than 1.5 GeV/c$^2$ ($\sigma =$~140~mb); 5,141 J/$\Psi$ events 
($\sigma \cdot Br.(e^+e^-) =$~1.9~mb); 
and 122 $\Psi'$ events  
($\sigma \cdot Br.(e^+e^-)$=~45~$\mu$b). 

The events were processed through AliRoot, the framework for the ALICE detector response simulation and 
off-line event reconstruction. The simulation gave about 3,500 reconstructed $\gamma \gamma \rightarrow e^+ e^-$ 
continuum pairs, 500 reconstructed J/$\Psi$s and 10 reconstructed $\Psi'$s. The vector meson decay 
products and the continuum $e^+e^-$--pairs have very different angular distributions and thus different 
acceptances. The simulation did not include the vertex reconstruction efficiency ($\approx$85~\% for 
two-track events) and the electron/positron identification efficiency. 

The invariant mass spectrum of the reconstructed $e^+e^-$--pairs is shown in Fig.~2. A clear J/$\Psi$ 
peak is visible above the continuum background. The 10 reconstructed $\Psi'$ cannot be distinguished 
from the background with the current statistics. 

\begin{table*}[tb]
\caption{Vector meson cross sections\cite{Klein:1999qj}, production and detection rates, and 
geometrical acceptances in ALICE. The calculations are for a nominal heavy-ion month of $10^6$ seconds 
at the luminosity $5 \cdot 10^{26}$~cm$^{-2}$s$^{-1}$.}
\label{table:1}
\newcommand{\m}{\hphantom{$-$}}
\newcommand{\cc}[1]{\multicolumn{1}{c}{#1}}
\renewcommand{\tabcolsep}{0.75pc} 
\renewcommand{\arraystretch}{1.2} 
\begin{tabular}{@{}lllllll}
\hline
Meson            & $\sigma(PbPb \rightarrow PbPb+V)$ & Prod. rate & Decay  & Br. ratio & Geo. Acc. & Det. rate \\  
\hline
$\rho^0$         & 5.2 b      & 2.6$\cdot10^9$ & $\pi^+ \pi^-$ & 100 \% & 0.079 & 2.0$\cdot10^8$ \\ 
J/$\Psi$         & 32 mb      & 1.6$\cdot10^7$ & $e^+ e^-$ & 5.93 \% & 0.101 & 1.0$\cdot10^5$ \\  
$\Upsilon$(1S)   & 280 $\mu$b & 1.5$\cdot10^5$ & $e^+ e^-$ & 2.38 \% & 0.141 & $\approx$400 \\ 
\hline 
\end{tabular}\\[2pt]
The geometrical acceptances for J/$\Psi$ and $\Upsilon$ are different compared with those in \cite{Alessandro:2006yt}. 
The earlier values were incorrect because of incorrect angular distributions of the decay products in the 
event generator. 
\end{table*}

As was mentioned above, the identification of the coherently produced vector mesons and two-photon 
events will rely on the low $p_T$ of the final state. The generated distribution is shown in 
Fig.~3 a). It is the result of a convolution of the photon $p_T$ spectrum with the form factor of 
the target nucleus. The corresponding distribution for the reconstructed events is shown in Fig.~3 b). 
Because of the finite momentum resolution, the reconstructed distribution is wider than the 
generated one, but it is still clearly peaked below $p_T <$~100~MeV/c. This shows that the momentum 
resolution is good enough for the summed $p_T$ to be used to identify the events. 

It should be noted that the $p_T$ in Fig.~3 is the $p_T$ of the final state (the vector meson 
or the two-photon system). The electrons/positrons have transverse momenta of about 
1.5~GeV/c.  


\section{Exclusive Vector Meson Production in proton-proton Collisions}

\begin{figure}[htb]
\vspace{9pt}
\begin{center}
\includegraphics*[width=6.5cm]{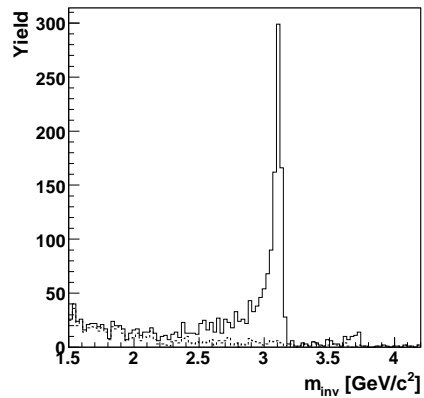}
\end{center}
\caption{Invariant mass distribution of $e^+e^-$--pairs in ultra-peripheral proton-proton collisions. The dashed 
histogram shows the continuum contribution from two-photon interactions.}
\label{fig:minvpp}
\end{figure}

\begin{figure*}[htb]
\vspace{9pt}
\begin{center}
\includegraphics*[width=5.6cm]{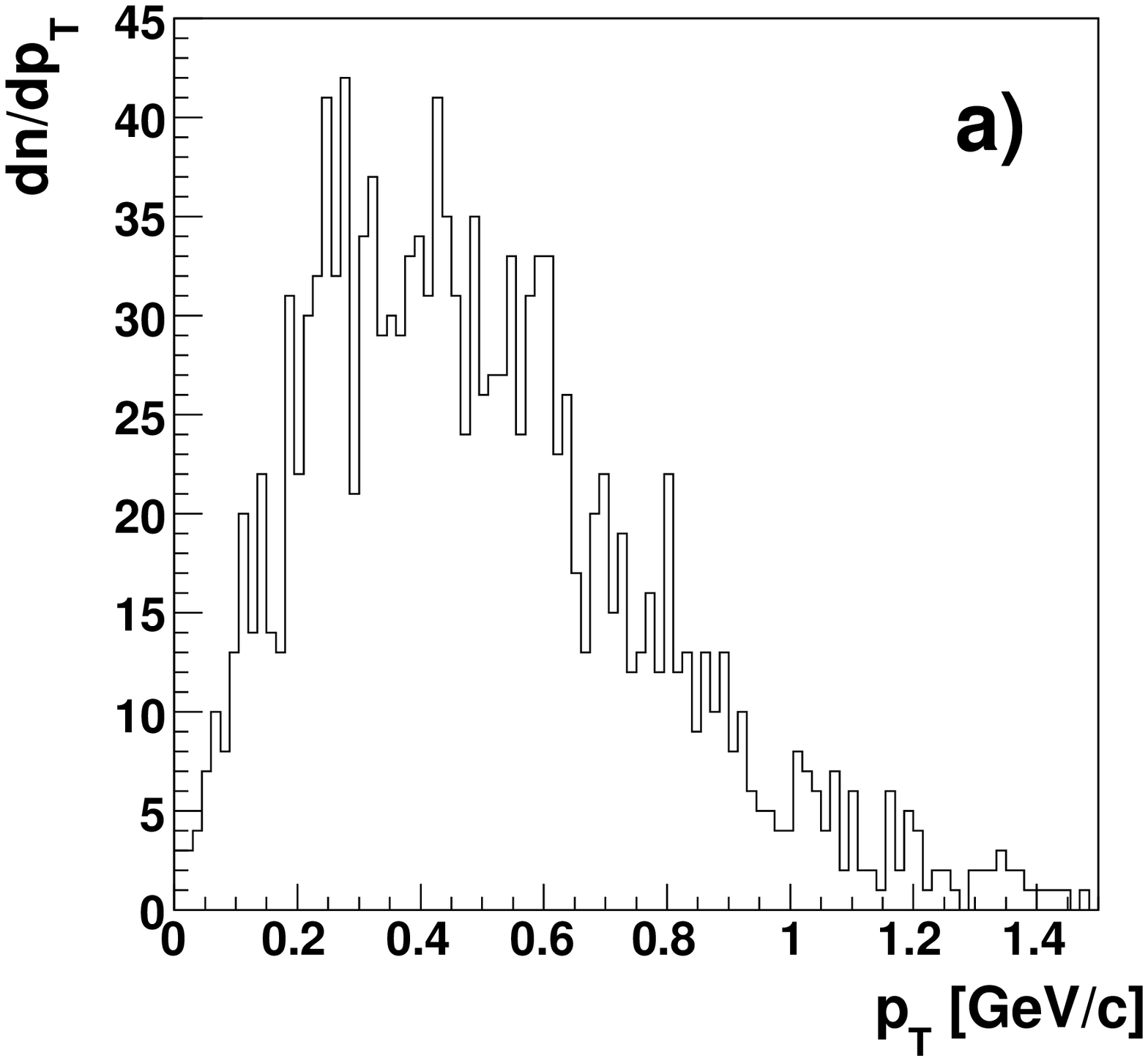}
\includegraphics*[width=5.6cm]{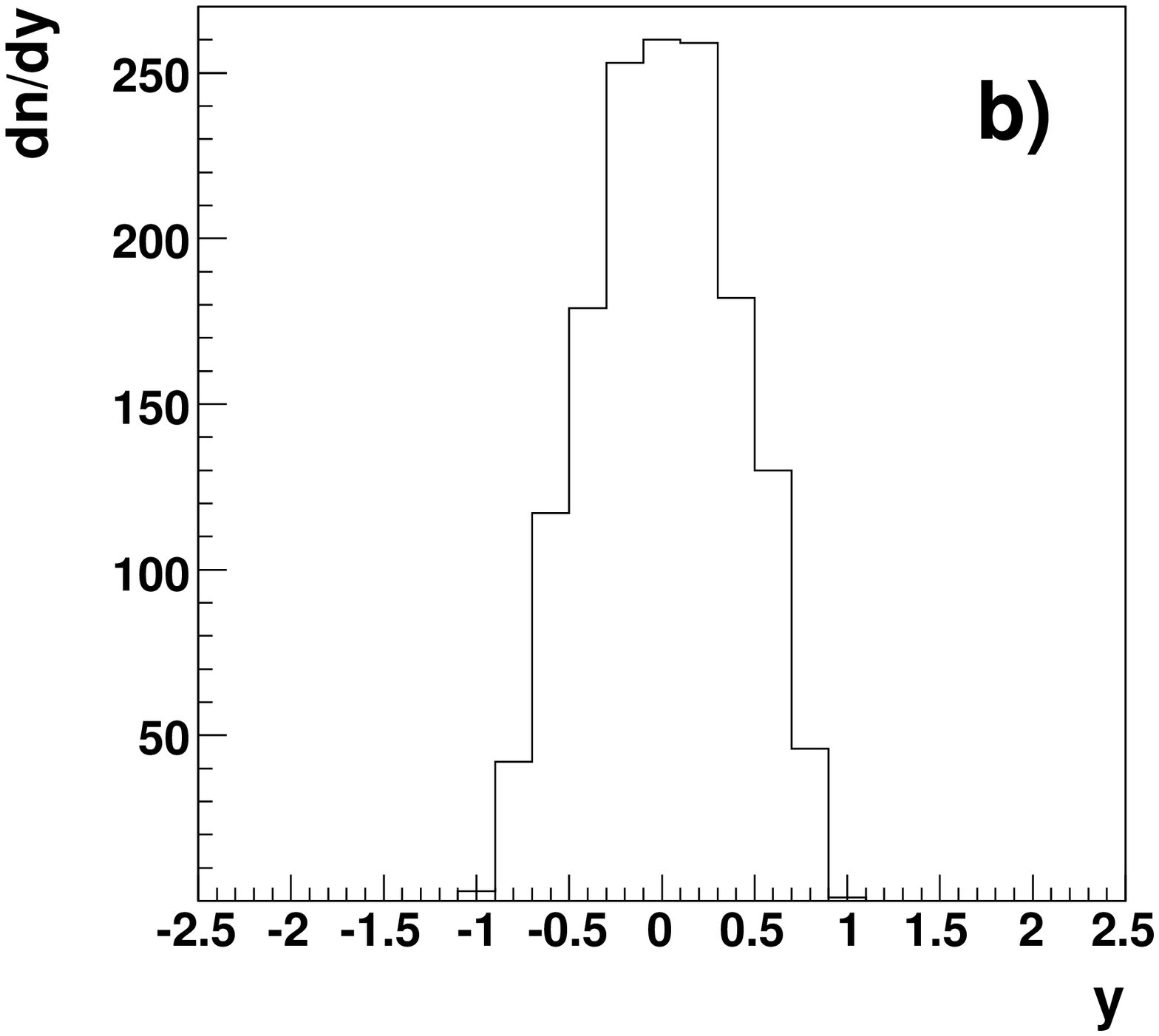}
\end{center}
\caption{Reconstructed transverse momentum (a) and rapidity (b) distributions of photoproduced J/$\Psi$ and 
$\Psi'$ in proton-proton collisions.}
\label{fig:yrapptpp}
\end{figure*}

Exclusive vector meson production can be studied also in proton-proton (pp) collisions. This reaction 
channel is interesting for several reasons. 
In pp collisions, Odderon-Pomeron fusion is a competing 
process by which vector mesons can be produced\cite{Schafer:1991na}. This could also occur in 
nucleus-nucleus collisions, 
but one expects photoproduction to dominate there. If the cross 
section for photoproduction can be accurately determined, for example by using data from HERA, any 
excess can be attributed to the Odderon. One furthermore expects the Odderon and photon to have 
different $p_T$ spectra. 
If the Odderon contribution can be separated, measuring 
exclusive production of $\Upsilon$ in pp collisions at the LHC could improve the statistics and the energy 
range of the measurements of exclusive photoproduction of $\Upsilon$ at HERA. 
Finally, it has been proposed to use two-photon production 
of di-lepton pairs for luminosity calibration in pp collisions. Photoproduction of vector mesons  
will then be an important background that must be understood. 

A simulation similar to that for Pb+Pb collisions has been performed based on about 18 days of running at 
the ALICE p+p luminosity $5 \cdot 10^{30}$~cm$^{-2}$s$^{-1}$. This luminosity is a factor $2 \cdot 10^3$ lower 
than the LHC design luminosity because of the long dead time of the TPC. 
Three samples of events were generated for pp collisions: 150,000 
$\gamma \gamma \rightarrow e^+ e^-$ events with an invariant mass of the $e^+e^-$--pairs 
greater than 1.5 GeV/c$^2$ ($\sigma =$~19~nb); 35,796 J/$\Psi$ events  
($\sigma \cdot Br.(e^+e^-) =$~4.5~nb); 
and 729 $\Psi'$ events ($\sigma \cdot Br.(e^+e^-) =$~91~pb). 

As for Pb+Pb, the pp events were processed through the ALICE detector response simulation 
and off-line event reconstruction. The simulation gave about 700 reconstructed 
$\gamma \gamma \rightarrow e^+ e^-$ continuum pairs, 1,400 reconstructed J/$\Psi$s and 30 
reconstructed $\Psi'$s. 
The invariant mass spectrum of the reconstructed $e^+e^-$--pairs is shown in Fig.~4. One notes 
that the continuum background from two-photon interactions is much smaller in pp 
compared with Pb+Pb collisions. The background is small enough for the peak from the 30 
reconstructed $\Psi'$s to be visible.

The reconstructed transverse momentum and rapidity distributions are shown in Fig.~5 a) and b), 
respectively. The $p_T$ spectrum is considerably wider than in heavy-ion collisions, and not very 
different from that of background processes, because of the different 
form factors. The identification of the exclusive production must therefore rely more on the presence
of rapidity gaps in pp collisions. 


\section{Summary and Conclusions} 

The rates for many interesting ultra-peripheral reaction channels are high within the acceptance 
of the ALICE detector. This study shows that with a suitable trigger  
these events can be reconstructed with high efficiency in ALICE. The momentum resolution is 
good enough for the summed $p_T$ to be used to identify coherent events in nucleus-nucleus collisions.

\end{document}